\documentclass[12pt]{article}
\usepackage{axodraw}
\usepackage{epsfig}   
\usepackage{amssymb}
\usepackage{latexsym}
\usepackage{amsmath}

\usepackage{graphics,subfigure} 

\usepackage{color}   

\newcommand{\lsim}{\raisebox{-0.13cm}{~\shortstack{$<$ \\[-0.07cm] $\sim$}}~}
\newcommand{\gsim}{\raisebox{-0.13cm}{~\shortstack{$>$ \\[-0.07cm] $\sim$}}~}
\newcommand{\bea}{\begin{eqnarray}}
\newcommand{\eea}{\end{eqnarray}}
\newcommand{\beq} {\begin{equation}}
\newcommand{\eeq} {\end{equation}}

\begin{document}
\pagestyle{empty}
\begin{flushright}
\today
\end{flushright}
\begin{center}
{\large\sc {\bf Thermodynamic properties of the Kerr Black-hole in non-linear 
	electrodynamics with cosmological constant 
}}
\end{center}
\vspace{1.0truecm}
\begin{center}
{\large Vinayak S. Pawar \footnote{Email: pvinayak616@gmail.com} and Siba Prasad Das\footnote{Email: spd.phy@unishivaji.ac.in}}\\
\vspace*{5mm}
{}$^1${\it  
Department of Physics, Faculty of Science, \\[0.15cm]
Dattajirao Kadam Arts, Science and Commerce College\\[0.15cm]
Survey No.17, 436, Kolhapur Road, Shivajinagar, Ichalkaranji, Maharashtra 416115, India} 
\\[0.07cm]
{}$^2${\it 
Department of Physics, Shivaji University Kolhapur,\\[0.15cm]
Vidya Nagar, Kolhapur, Maharashtra 416004, India} \\[1.0cm]
\end{center}

\begin{abstract}
        We study thermodynamic properties, in particular the Temperature~(T), Angular velocity~($\Omega_h$) 
	and Entropy~(S) of the of magnetically charged slowly rotating (with rotation parameter $a \lsim 0.10$) 
	Kerr black-hole(BH) with the inclusion of cosmological constant ($\Lambda$) in the background of nonlinear 
	electrodynamics (NLED). At first we calculated the nonlinear electromagnetic magnetic charged 
	density $\rho_{NLED}$$(r)$ which is needed to calculate the magnetic mass of the slowly rotating 
	Kerr-BH. We showed the mass profile $M(r)$ of the BH for different combinations of magnetic 
	charges~($q_m$) and non-linear parameter ($\beta$) presence in the Lagrangian density. 
	We found that $M(r)$ attains a plateau for values 
	of $r$ close to the the cosmological length~($L$), where $L^2$= $\pm \frac{\Lambda}{3}$, 
	irrespective of the combinations of $q_m$ and $\beta$. The $\pm$ sign corresponds to the de-Sitter(dS) 
	and Anti-de-Sitter(AdS) respectively. Afterwards we showed the allowed parameter spaces in  
	$a$-$M$ plane using sharkfin diagram for different values of $L$ with positive 
	values of the horizon function, $\Delta(r)$, and explain the extremal criterion and 
	asymptotic limit. We showed the values of $r$ where the  horizon function  
	of the quadratic polynomials becomes zero and called them as the inner(Cauchy), outer(Event) 
	and large cosmological horizons with different values of $a$. We showed that the horizon structure 
	depends on $a$, $L$ and the mass profile $M(r)$. Finally, we tabulated the numerical values 
	of three thermodynamic parameter, i.e., $T$, $\Omega_h$ and $S$ at those horizons surfaces.  
	Our results demonstrate that NLED with cosmological constant significantly 
	modifies both the internal structure and thermodynamic properties of Kerr-BH.
\end{abstract}

\newpage
\setcounter{page}{1}
\pagestyle{plain}

\section{Introduction}
\label{sec:intro}

	 Black holes(BH) emerged as one of the most intriguing results of general theory of relativity (GTR). 
It describes a region where the curvature of space-time is so intense that nothing, not even light, can escape. 
Generally the BHs are characterized by their radius, masses, charges, angular momentum etc. 
Depending upon the parameter choices BH could have different types, e.g. the Schwarzschild, 
Kerr, Kerr-Newman, Reissner Nordstr\"{o}m\cite{Jacobson:2012ei, Curiel:2018cbt}. 
In our analysis we considered the Kerr solutions, i.e. the BH with rotation parameters ($a$).
The evolution of the different composite stars estimates that the rotation parameters could  
have varied ranges of 0.01 to 0.60 with different characteristics features \cite{Miller:2014aaa}. 
The slowly rotating Kerr-BH in our consideration is with values of spin parameter $a$ is 0.1 or less.

	It has long been studied in the literature \cite{Bardeen,Jacobson,Padmanabhan} that 
BH can be treated as a thermodynamics system. The areas of the BH is associated with the entropy 
while the surface gravity corresponds to the temperature \cite{Bekenstein,Hawking}. 
These results provide a deep connection between Gravitation, Thermodynamics, and Quantum theory 
-- all these are the ingredient of the bigger picture of the theory of Quantum Gravity.

	 The cosmological constant ($\Lambda$) is a term in general relativity that represents the energy 
density of empty space (vacuum energy) that plays a crucial role in the dynamics of the Universe. It 
acts as a repulsive force (against gravitational attraction) driving the accelerated expansion 
of the Universe. The classical theory suggest that the $\Lambda$ as a natural candidate for 
dark energy. The theory of Quantum Gravity predicts $\Lambda$ is extremely large vacuum energy. 
However experimental observations show it's extremely small but non-zero and these dis-parity 
in the magnitude is called the ``cosmological constant problem''.

	 The values of the $\Lambda$ could be either positive or negative for the de-Sitter(dS) and 
	 Anti-de-Sitter(AdS) respectively. The study of black holes in AdS background
has revealed rich thermodynamic behavior. For example, the existence of phase transitions 
in AdS-BH was first demonstrated in~\cite{HawkingPage1983} and for charged black holes 
in~\cite{Chamblin1999,KubiznakMann2012}. Moreover, gravity in AdS background provides a natural 
framework for holography through the AdS/CFT correspondence~\cite{Maldacena1998, Witten1998, Gubser1998}. 
In this picture a gravitational theory in the bulk is dual to a conformal field theory (CFT) on the boundary. 
This duality has found wide-ranging applications, e.g., system of strongly coupled quantum 
chromodynamics (QCD)~\cite{Casalderrey2014}, condensed matter physics~\cite{Hartnoll2009} and etc.

	Nonlinear electrodynamics (NLED) has attracted significant attention in astrophysics 
\cite{Sarkar:2022jgn,Escobar:2021mpx,Mignemi:2021djm,kzz,Giri:2021wxu}. NLED could have 
combined with GTR and this combined framework exhibits several notable features, including the 
ability to provide a consistent description of inflationary models of the Universe 
~\cite{Garcia,Camara,Elizalde,Novello,Novello1,Vollick,Kruglov3,Kruglov31,Kruglov32, 
kruglov1,Kr0,Kr2,Bronnikov}. Moreover, this framework explains the absence of 
initial singularities, imposes an upper bound on the electric field at the origin of point-like 
particles and ensures a finite self-energy for charged particles~\cite{Novello1, Kr0,Born}. 
It is to be noted that in quantum electrodynamics (QED) the non-linear terms arises due to  
the loop corrections \cite{Heisenberg, Schwinger, Adler}. As usual the NLED-GTR model 
follows the correspondence principle, i.e., in the weak-field limit we get the linear electrodynamics.

	The physics of BH (both electrically and magnetically charged) in presence of NLED has been studied  
extensively~\cite{kruglov1,Bronnikov,Bardeen2,Dymnikova,Beato,Breton,Hayward,Lemos,Flachi,Hendi,Balart,SIKruglov,SIKruglov1,SIKruglov2,Das2023}. In our analysis first we have taken into the effect of the dS   
length to estimate the magnetized mass of the Kerr-BH. Secondly, we have shown how the Kerr-BH horizons 
change in presence of the cosmological constant with different spin-parameter $a$. Finally we 
embedded this two approaches to estimate the entropy and temperature at the horizons of the slowly 
rotating Kerr-BH in NLED-(A)dS model. To the best of our knowledge, the  thermodynamic properties of 
slowly rotating Kerr-BH in NLED model with cosmological constant have not been systematically 
investigated. In this work we aim to fill this gap.

	The paper is organized as follows: we briefly outline the NLED-AdS model in section 2. In section 3 we 
discussed the Kerr-BH with taking into account the effect of cosmological constant and 
called as NLED-(A)dS model. We have also calculated the magnetized mass density. We 
demarcated the allowed regions of the NLED-(A)dS 
model using the sharkfin diagram and estimated the 
length of the horizon radius in section 4. Finally, we tabulated tree relevant thermodynamic parameters, 
temperature, angular velocity and entropy at the vicinity of the horizons. We 
conclude in Section 5. We adopted natural units, i.e., $c=\hbar=1$, 
$\varepsilon_0=\mu_0=1$ and with metric signature $\eta=\mbox{diag}(-1,1,1,1)$

\section{The NLED-AdS model}
\label{sec:nledmodel}

In GTR the action of NLED-AdS theory is defined as:
\begin{equation}
I=\int d^{4}x\sqrt{-g}\left(\frac{R-2\Lambda}{16\pi G}+\mathcal{L}(\mathcal{F}) \right),
\label{eqn1}
\end{equation}
where $G$ is the Newton constant. The cosmological constant, $\Lambda$, is negative (attractive)  and 
given by  $\Lambda=-3/L^2$ where the AdS{\footnote{For dS the sign 
is reversed, i.e., $\Lambda=+3/L^2$ is in repulsive in nature in contrary to normal gravitational attraction.}} 
length is $L$.

We follow~\cite{kruglov1,Kr0,Kr2} and consider the NLED-Lagrangian is given in ~\cite{kruglov1}:
\begin{equation}
{\cal L}(\mathcal{F}) =-\frac{{\cal F}}{1+\sqrt{2|\beta{\cal F}|}}.
\label{eqn2}
\end{equation}
The field-strength ${\cal F}$ is Lorentz invariant and given by ${\cal F}=F^{\mu\nu}F_{\mu\nu}/4=(B^2-E^2)/2$,
where $E$($B$) is the electric (magnetic) induction field. In the limit of $\beta=0$ (vanishing non-linearity),
one can have linear Maxwell electrodynamical AdS BHs.

One can obtain the gravitation and electromagnetic field equations from the action of Eqn.\ref{eqn1}:
\begin{equation}
R_{\mu\nu}-\frac{1}{2}g_{\mu \nu}R+\Lambda g_{\mu \nu} =8\pi G T_{\mu \nu},
\label{eqn3}
\end{equation}

\begin{equation}
\partial _{\mu }\left( \sqrt{-g}\mathcal{L}_{\mathcal{F}}F^{\mu \nu}\right)=0,
\label{eqn4}
\end{equation}

where $\mathcal{L}_{\mathcal{F}}=\partial \mathcal{L}( \mathcal{F})/\partial \mathcal{F}$. The
electromagnetic fields stress tensor is given by

\begin{equation}
T_{\mu\nu }=F_{\mu\rho }F_{\nu }^{~\rho }\mathcal{L}_{\mathcal{F}}+g_{\mu \nu }\mathcal{L}\left( \mathcal{F}\right).
\label{eqn5}
\end{equation}

We are working in {\it {nearly}} spherical symmetric limit, as in our considerations the rotation parameter of 
the BHs are very small ($a \leq 0.1$). In this limit the metric tensor is given by \cite{Bronnikov}:
\begin{equation}
ds^{2}=-f(r)dt^{2}+\frac{1}{f(r)}dr^{2}+r^{2}\left( d\theta
^{2}+\sin ^{2}\theta d\phi ^{2}\right).
\label{eqn6}
\end{equation}

The stress-energy tensor $F_{\mu\nu}$ possesses the radial electric field $F_{01}=-F_{10}$ and radial
magnetic field $F_{23}=-F_{32}=q_m\sin\theta$, where $q_m$ is the magnetic charge $q_m$.
Moreover, the stress energy tensor is diagonal, i.e., $T_{0}^{~0}=T_{r}^{~r}$ and
$T_{\theta}^{~\theta}=T_{\phi}^{~\phi}$.

The metric function $f(r)$ in Eqn.\ref{eqn6} is given by
\begin{equation}
f(r)=1-\frac{2m(r)G}{r},
\label{eqn7}
\end{equation}

and the mass function is given by

\begin{equation}
M(r)=m_0+\int_{0}^{r}\rho (r)r^{2}dr.
\label{eqn8}
\end{equation}

In Eqn.\ref{eqn8}, the integration constant ($m_0$) is considered to be the static Schwarzschild mass 
(resides in the center of the BH) and $\rho(r)$ is the energy density that contributes to the total 
mass after integration is performed. It has been studied in \cite{Bronnikov} that models with only 
electrically charged BHs having Maxwell's weak-field limit possess singularities. Thus in our analysis 
we consider electric (magnetic) charge of the BHs is $q_e$=0 ($q_m \neq 0$). The field strength of 
pure magnetically charged BHs is $\mathcal{F}=q_m^2/(2r^4)$.

The energy density with the negative cosmological constant is given by
\begin{equation}
	\rho= \rho_{NLED} + \rho_{\Lambda} = \frac{q_m^2}{2r^2(r^2+q_m\sqrt{\beta})}-\frac{3}{2G L^2}, 
\label{eqn9}
\end{equation}

where $\rho_{NLED}$ and $\rho_{\Lambda}$ are the charge density due to NLED and cosmological 
contribution respectively. The $\rho_{NLED}$ is a converging  parameter and gives a finite 
mass contribution to the total mass of the black hole once the integration is performed. 
On contrary the cosmological charge density $\rho_{\Lambda}$ is diverging. Hence we have not taken 
into consideration of this diverging term while performing the total integration 
to calculate the  mass of the BHs. 

From Eqs.\ref{eqn8} and \ref{eqn9} one can obtain the mass profile as a function of the
radial co-ordinates as:
\begin{equation}
M(r)=m_0+\frac{q_m^{3/2}}{2\beta^{1/4}}\arctan\left(\frac{r}{\sqrt{q_m}\beta^{1/4}}\right)-\frac{r^3}{2G L^2}.
\label{eqn10}
\end{equation}

\begin{equation}
M(r)=m_0+\frac{q_m^{3/2}}{2\beta^{1/4}}\arctan\left(\frac{r}{\sqrt{q_m}\beta^{1/4}}\right).
\label{eqn10b}
\end{equation}


\section{Kerr metric in Boyer-Lindquist(BL) coordinates}
\label{sec:blcoordinate}

The Kerr metric in Boyer-Lindquist coordinates with spin parameter~($a$) is:
\[
\begin{aligned}
ds^2 = &-\frac{\Delta_r}{\rho^2} \left( dt - \frac{a \sin^2\theta}{\Xi} d\phi \right)^2 + \frac{\rho^2}{\Delta_r} dr^2 + \frac{\rho^2}{\Delta_\theta} d\theta^2 \\
&+ \frac{\Delta_\theta \sin^2\theta}{\rho^2} \left( a dt - \frac{r^2 + a^2}{\Xi} d\phi \right)^2
\end{aligned}
\]
where:
\[
\begin{aligned}
\rho^2 &= r^2 + a^2 \cos^2\theta \\
\Delta_r &= (r^2 + a^2)\left(1 - \frac{\Lambda}{3} r^2\right) - 2Mr \\
\Delta_\theta &= 1 + \frac{a^2 \Lambda}{3} \cos^2\theta \\
\Xi &= 1 + \frac{a^2 \Lambda}{3}
\end{aligned}
\]

\subsection{Entropy, Hawking temperature and Angular velocity}
\label{sec:entropy}

For the Kerr-(A)dS background, the entropy is given by the Bekenstein--Hawking law,
\begin{equation}
S = \frac{\pi \left(r_h^2 + a^2\right)}{\Xi}.
\label{entropy}
\end{equation}
This expression shows that rotation modifies 
the effective horizon area while the cosmological constant in the denominator 
introduces new scaling.

The Hawking temperature is determined from the surface gravity $\kappa$ at the event horizon,
\begin{equation}
T = \frac{\kappa}{2\pi}.
\label{temp}
\end{equation}
For the Kerr--(A)dS background, the surface gravity is obtained from the derivative of the 
radial function $\Delta_r$ and taking the mod values at the horizons, as follows: 
\begin{equation}
\kappa = \frac{1}{2(r_h^2 + a^2)} \left. \frac{d\Delta_r}{dr} \right|_{r=r_h}.
\label{kap}
\end{equation}

The angular velocity at the horizon is given by
\begin{equation}
\Omega_h = \frac{a \, \Xi}{r_h^2 + a^2}= \frac{a \, \pi}{S}.
\label{angular}
\end{equation}
This quantity characterizes the rotational frame dragging at the horizon and 
important while calculating the first law of BH thermodynamics 
for rotating Kerr-BH scenario.

\section{Numerical Analysis}
\subsection{de-Sitter horizons}
\label{sec:dsh}

The dS horizon function $\Delta_{dS}(r)$ for the cosmological background is the 
following form $\Delta_{dS}(r) = (r^2 + a^2)\left(1 - \frac{r^2}{L^2}\right) - 2Mr$. 
The real solutions for the event horizons obtained from the following 
critical conditions: $\Delta_{dS}(r)=0$ and $\frac{d\Delta_{dS}(r)}{dr}=0$.
After simplification, the necessary and sufficient condition for having 
the real horizon is: $L^2 \gsim a^2 + 27 M^2$. In the Schwarzschild limit $L \gsim 3 \sqrt 3 M$. 
The horizons exist if $a \lsim M $ is called the standard Kerr bound. The horizons depends upon 
the following relations: (i) if $L^2 \lsim a^2 + 27 M^2$ then there are no real horizons 
(ii) if $L^2 = a^2 + 27 M^2$ this leads to degenerate roots (extremal condition) and (iii) 
if $L^2 \gsim a^2 + 27 M^2$ then there are three distinct real horizons.

The critical values of the dS length obtained from the condition for the real 
solutions for the event horizons, which is given as, $L_{cr} = \sqrt {a^2 + 27 M^2}$. In our analysis 
we have considered the slowly rotating BHs with $M=\frac{\pi}{4}$ and $a\lsim 0.1$ this gives 
the values of $L_{cr} =4.08$. This critical length signifies a fundamental geometric threshold which shows the 
equilibrium point between gravitational attraction of the BH and the repulsive effect of the 
dS cosmological background. When the dS length falls below $L_{cr}$ the 
cosmological expansion dominates and prevents the formation of distinct event horizons. 
At the critical limit, all the three horizons -- the inner, event and cosmological are merged and 
is called the extremal configuration. 

In absence of the cubic term in the expression of a quartic polynomial satisfies the 
Viete's relation, i.e., the sum of all four solutions must be zero. This reflects the 
global geometric balance imposed by the dS background. Out of the four real roots, 
one of the root is negative and therefore unphysical -- the other three roots are called 
as the inner(Cauchy), outer(Event) and cosmological horizons. It is important to note that 
the presence of Cauchy horizon is due to the spin parameter of the black hole.

The Discriminant $\cal D$$(r,a,L,M)$ for the horizon function $\Delta(r)$ is the following:

\[
\begin{aligned}
        \cal D (r,a,L,M) = &-\frac{16}{L^{10}} (a^{10} + 4 a^8L^2 + 6 a^6L^4 + 4 a^4L^6 + a^2L^8 \\
&+ a^6L^2M^2 + 33 a^4L^4M^2  - 33 a^2L^6M^2 - L^8M^2 + 27 L^6M^4 )
\end{aligned}
\]

The masses of the BH discussed in the model above is{\footnote{Please note that while calculating the massless of the BH we have taken into consideration the contribution from  $\rho_{NLED}$ only.}}:

\begin{equation}
\begin{aligned}
	M(r) =\int_{0}^{r} \rho_{NLED}({r^{'}}) {r^{'}}^2 d{r^{'}}
\label{eqn11}
\end{aligned}
\end{equation}

Let us for the moment discuss about the mass distribution of the BHs solely due to the effect
of NLED. For this the parameters of relevances are only two: the magnetic
charge~($q_m$) and non-linear parameter ($\beta$).

$M(r)$ is stands for the mass contribution up-to the limit of the radial co-ordinate $r$.

\begin{equation}
        M(r)= \frac{{q_m}^{\frac{3}{2}}}{2 {\beta}^{\frac{1}{4}}} tan^{-1} \Biggl ( \frac{r}{{{q_m}^{\frac{1}{2}}} {\beta}^{\frac{1}{4}}} \Biggr )
\label{eqn12}
\end{equation}

\begin{equation}
	M(\infty)= {\frac{1}{4}} {\pi {q_m}^2 \Biggl ( \frac{1}{ {{q_m}^{\frac{1}{2}}} {\beta}^{\frac{1}{4}}}} \Biggr )
\label{eqn13}
\end{equation}

The above solutions of Eqns.\ref{eqn12},\ref{eqn13} are possible {\it {iff}}, $Img(q_m \beta^{\frac{1}{2}}) \neq 0$ and  $Re(q_m \beta^{\frac{1}{2}}) \geq 0$.

\begin{figure}[ht!]
\begin{center}
\raisebox{0.0cm}{\hbox{\includegraphics[angle=0,scale=0.24]{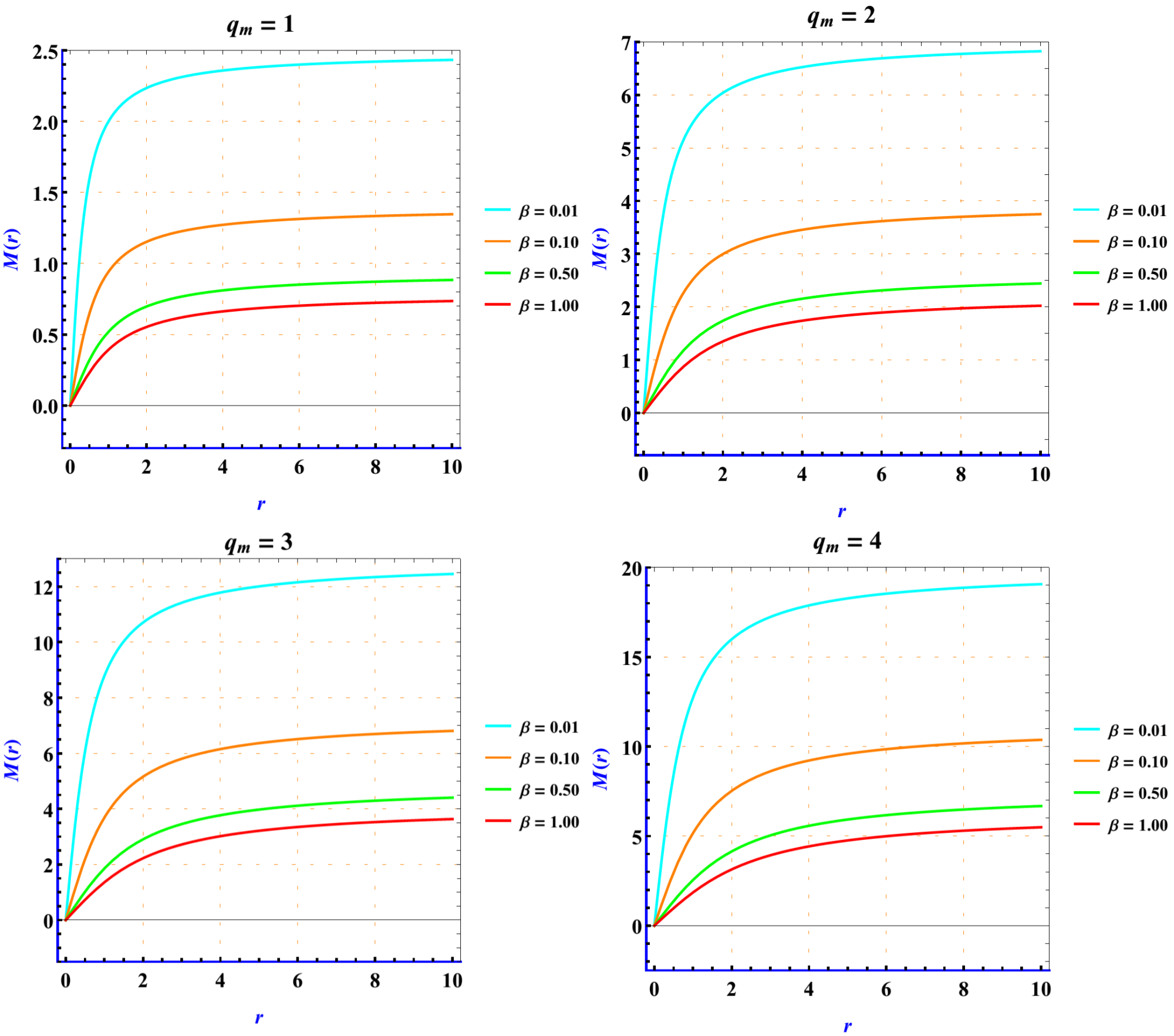}}}
\caption{The mass distribution of the Kerr-BH for different values of $\beta$ for $q_m$ = 1,2,3 and 4.}
\label{massdS1}
\end{center}
\end{figure}

In Fig.~\ref{massdS1} we have shown the mass distribution $M(r)$, i.e., Masses enclosed within 
the radius $r$ for regular extended source of the Kerr-BH as function of $r$ for different values 
of $\beta$ for $q_m$ = 1 ,2, 3 and 4. The mass 
profiles for all the figures with different values of $\beta$, $q_m$ are somewhat similar as a 
function of the distance. As we see that for all combinations of $\beta$ and $q_m$ from $r:[0-2]$, 
the mass builds up gradually following Eqn.\ref{eqn9} -- this is a hallmark of NLED model. The mass profiles 
have somewhat sharp-rise and for $r \gsim 5$ the masses are saturated and reaches a plateau. The total 
mass becomes finite with field energy contribution converges. It is also seen that for fixed values 
of $q_m$, say for $q_m=1$, when the values of $\beta$ is decreases from $1$ to $0.001$, for the same 
distance the total masses is increases from $M(r)=0.5$ to $2.5$. The reasons are the 
following: $\rho_{NLED} \propto q_m^2$, thus large magnetic charge leads to large field energy so 
the mass accumulation is more. On the other hand $\rho_{NLED} \propto \frac{1}{r^2 + q_m\sqrt \beta}$, thus 
increasing $\beta$ makes the denominator large which eventually make the energy density 
smaller and more diffuse distribution of the masses.

\begin{figure}[ht!]
\begin{center}
\raisebox{0.0cm}{\hbox{\includegraphics[angle=0,scale=0.24]{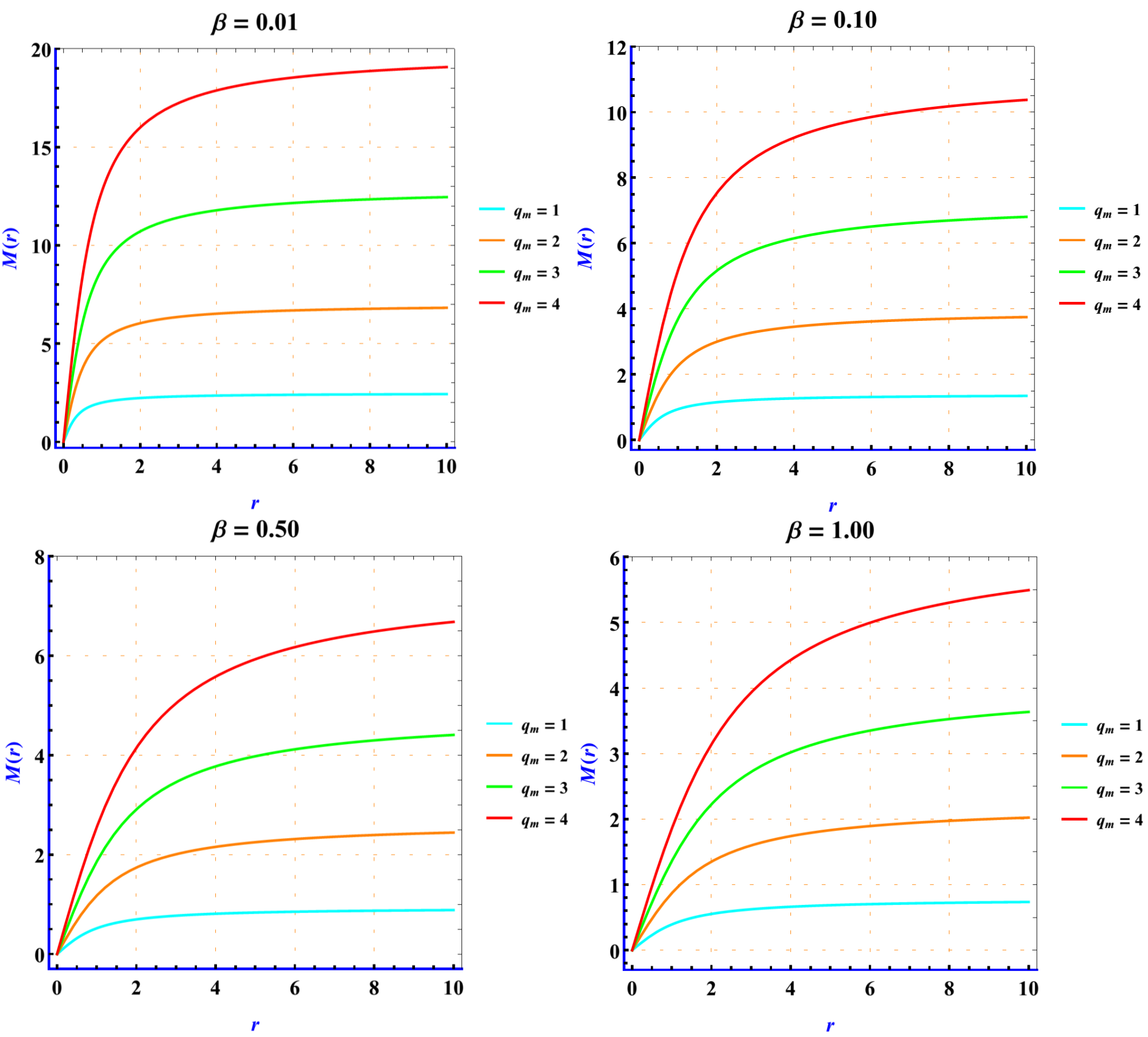}}}
\caption{The mass distribution of the Kerr-BH for different values of $q_m$ for
	$\beta$ = $0.01,0.10,0.50$ and $1.00$.}
\label{massdS2}
\end{center}
\end{figure}

In Fig.~\ref{massdS2} we have shown the same mass distribution $M(r)$ in complementary 
approaches to Fig.~\ref{massdS1}. In each figures the values of $\beta$ is fixed 
while varying the magnetic charges as $q_m$=1,2,3,4. The chosen values of $\beta$ are 0.01, 
0.10, 0.50 and 1.00. It is clear from the each figure that 
for a fixed values of $\beta$ the mass profiles reaches a plateau in much shorter distances of 
$r$ with low values of $q_m$.  As $\rho_{NLED}\propto q_m^2$, for larger $q_m$ the 
mass saturation happens much larger values of $r$. The overall $M(r)$ is much smaller when the values of $\beta$ is much larger as in the expression of $\rho_{NLED}$ the $\beta$ appears in the denominator.

The $M(r)$ with different values of $\beta$ reveal the role of the nonlinearity parameter  
in shaping the internal mass distribution. For smaller values of $\beta$ the mass profile $M(r)$   
grows more rapidly with $r$ which reflects the larger concentration of energy density 
near the central region. In contrast, for larger values of $\beta$ lead to a more 
gradual increase of $M(r)$ due to the suppression of the energy density consistent with   
the regularizing nature NLED.

\begin{figure}[ht!]
\begin{center}
\raisebox{0.0cm}{\hbox{\includegraphics[angle=0,scale=0.42]{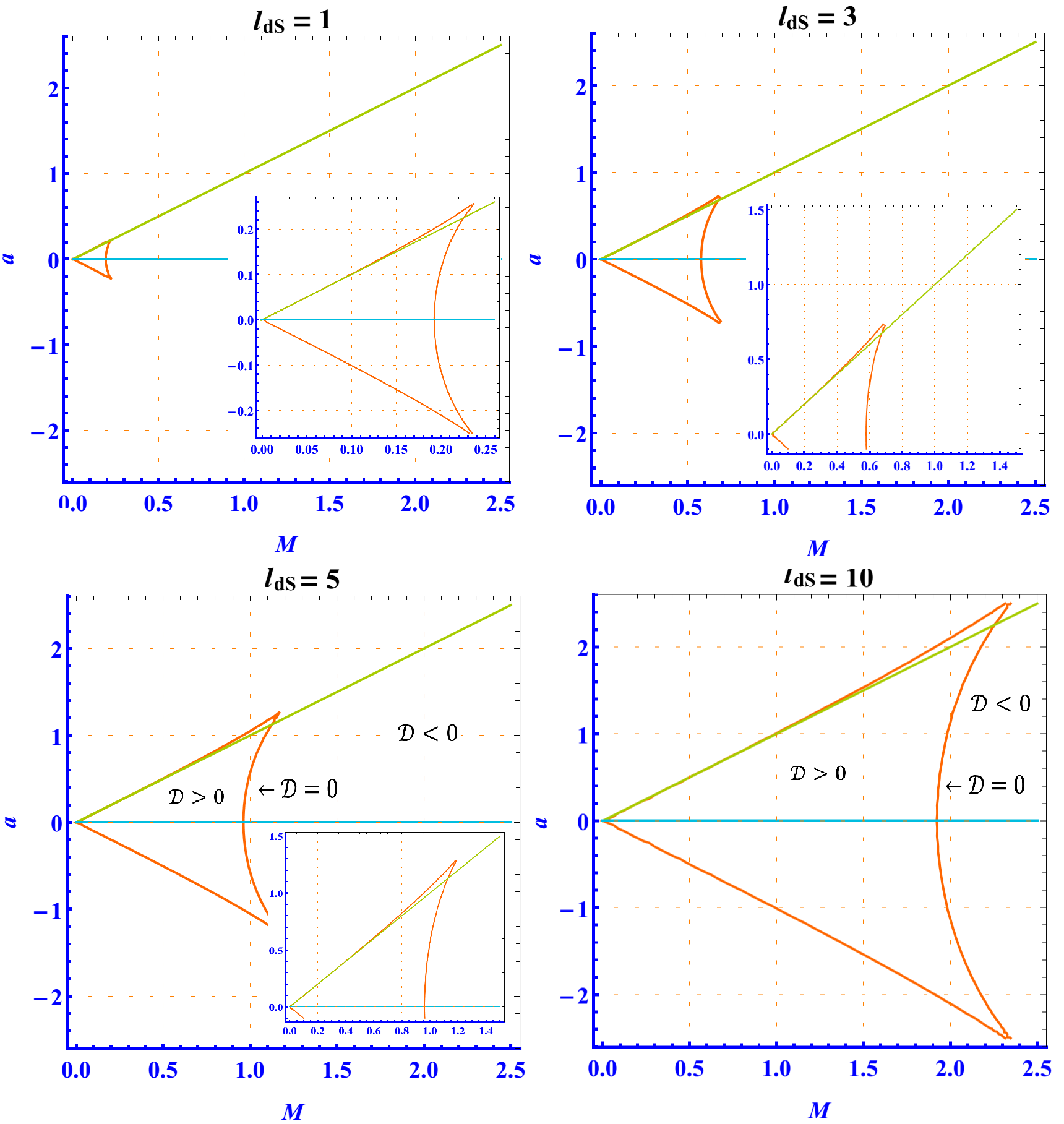}}}
\caption{The sharkfin diagram for spin parameter and the mass of the Kerr-BH for different values
	of the de-Sitter lengths $L$= $1,3,5$ and $10$.}
\label{sharkdS}
\end{center}
\end{figure}

In Fig.\ref{sharkdS} we have shown the allowed region for existence of real horizons 
in the $a-M$ plane for different $L_{dS}$ using sharkfin diagram. It is interesting to 
note that for $L= \infty$ the contribution from the dS length is vanishes. The sharkfin diagram of the 
bottom-right panel of Fig.\ref{sharkdS} is coincide with the extremal ($a=M$) case 
of the Kerr-BH. Thus with $a=0$ with $L_{dS}=\infty$ the BH goes to the 
limit of Schwarzschild case. This is clear from the bottom-panel of Fig.\ref{sharkdS} that 
out of three solutions of the horizon function $\Delta(r)=0$, one is the exact ($r=2M$) 
Schwarzschild radius. The bounded regions in the $a-M$ plane corresponds to $\cal D \gsim$$0$ 
with the existence of physically admissible horizons. The points inside the region correspond 
to non-extremal black holes with distinct inner (Cauchy), event and cosmological horizons. With the 
increased values of $L_{dS}$, the allowed region enlarges -- indicating that the large 
values of cosmological lengths weakens the repulsive nature of the dS background 
and allows more wider ranges of mass and spin values to form the BH.

\begin{figure}[ht!]
\begin{center}
\raisebox{0.0cm}{\hbox{\includegraphics[angle=0,scale=0.24]{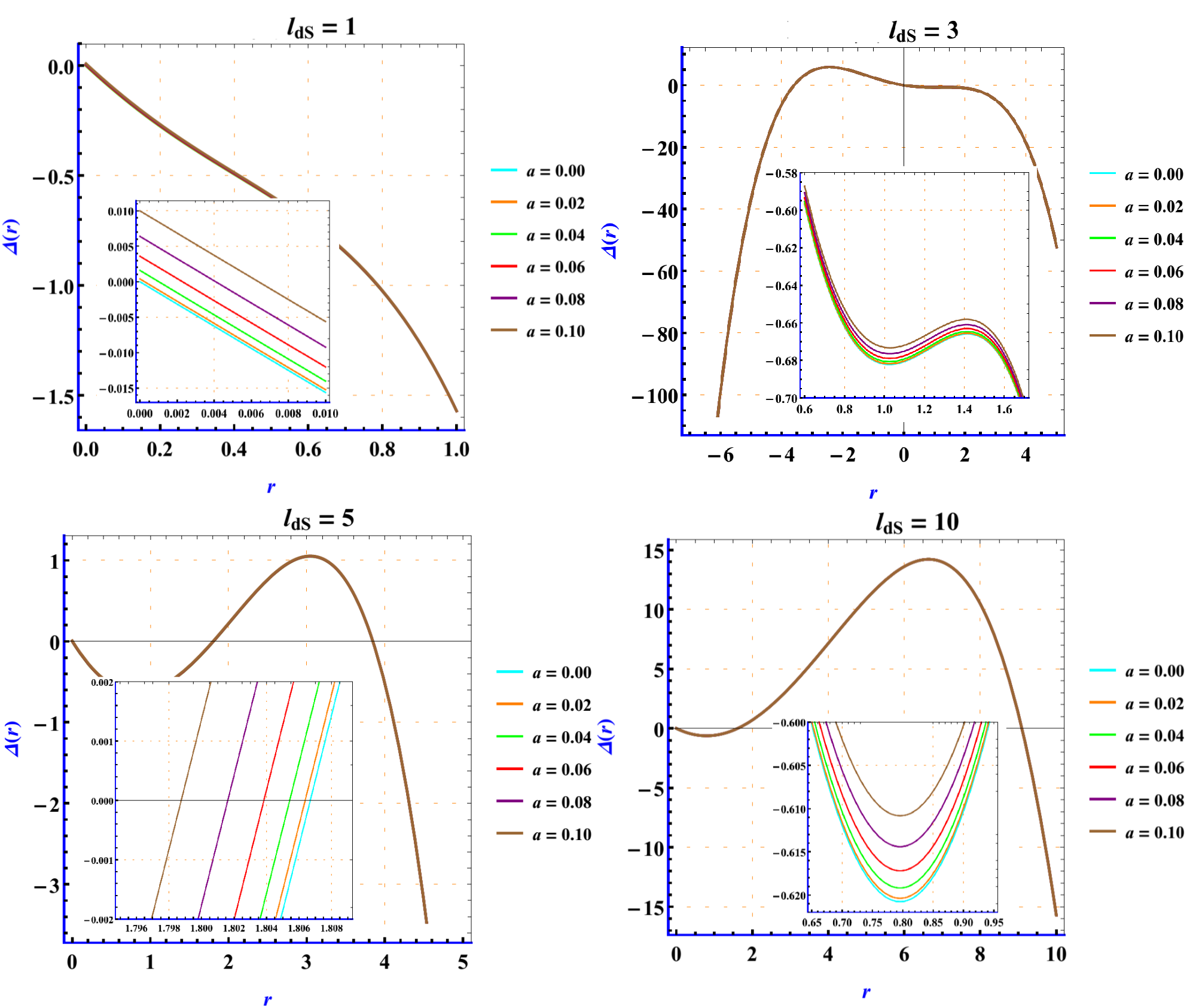}}}
\caption{
The variation of the discriminant $\Delta(r)$ term with radial distance ($r$) for 
different values of de-Sitter lengths and spin parameter of the Kerr-BH($a$ = $0.00$ to $0.10$). 
}
\label{discridS}
\end{center}
\end{figure}

In Fig.\ref{discridS} we have shown the horizon function $\Delta (r)$ profile as a function of $r$ for different  
values of rotation parameter $a$ and dS length $L_{dS}$. The horizon radii are the places where the 
values of $\Delta (r)=0$. We have understood that $\Delta (r)=0$ occurs with the minimum values of 
$L_{dS}$=4.08 which we called before as the critical length. This reflects from the top-panel and 
bottom-panel of Fig.\ref{discridS} -- for example for the top-panel there exits no real root exists 
as $L_{dS} \lsim $4.08 and from the bottom-panel it shows three distinct roots (Cauchy, event, and cosmological horizons) as  $L_{dS} \gsim $4.08. We have also varied the values of the $a$ parameter and seen that the 
horizons positions have been changes very mildly with non-zero $a$. To see the effect visibly we have 
also zoomed the regions of interest in the sub-figures and understood that with larger $a$, the gap 
between event and cosmological horizons increases. This increases are happening due to shift of 
event horizon into the left and cosmological horizons into right in the radial co-ordinate.

\subsection{Anti-de-Sitter horizons}
\label{sec:adsh}

\begin{figure}[ht!]
\begin{center}
\raisebox{0.0cm}{\hbox{\includegraphics[angle=0,scale=0.42]{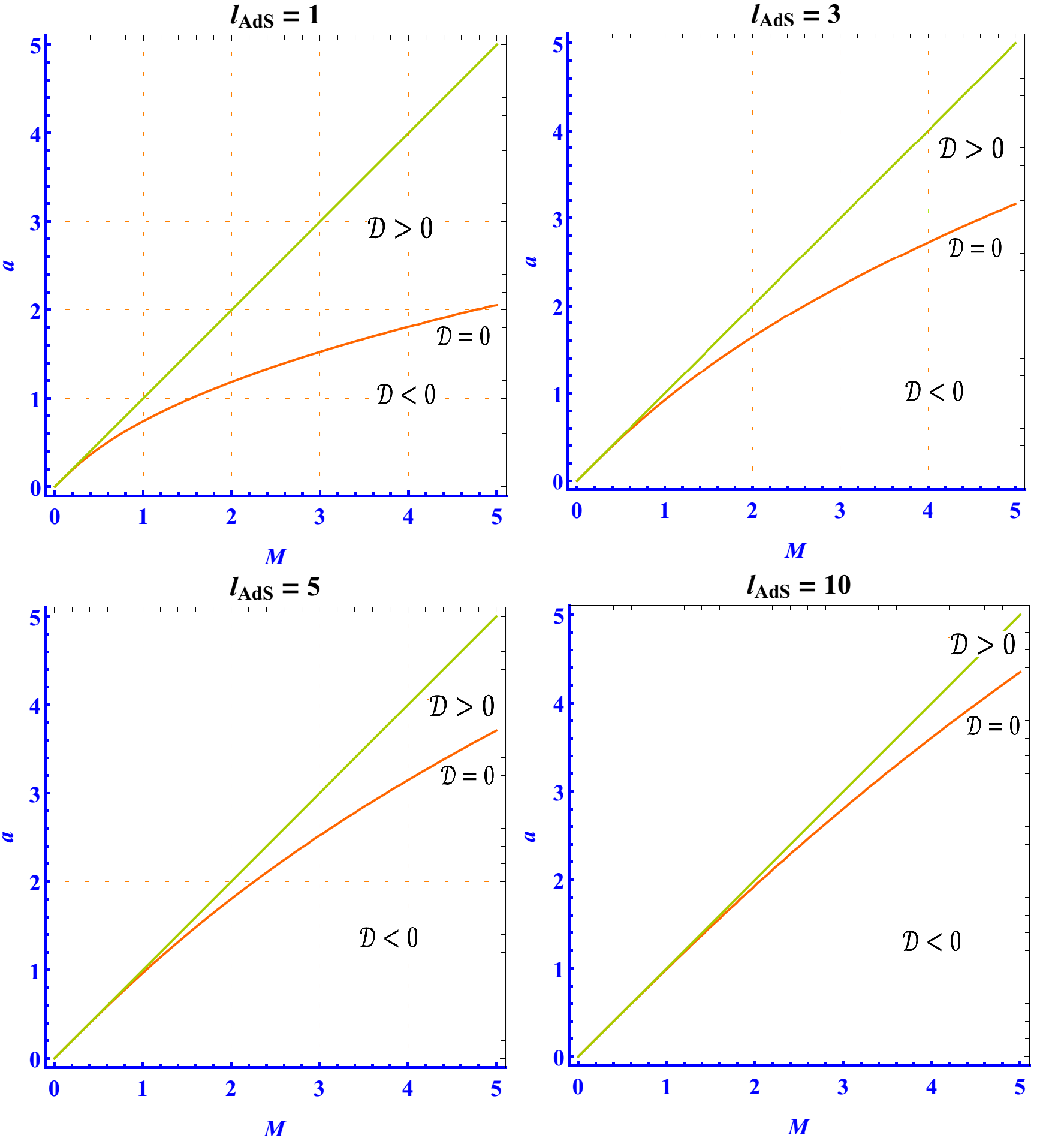}}}
\caption{ The sharkfin diagram for spin parameter and the mass of the Kerr-BH for different values
	of the Anti-de-Sitter lengths $L_{AdS} = 1,3,5$ and $10$.}
\label{sharkAdS}
\end{center}
\end{figure}

We have shown in Fig.\ref{sharkAdS} the allowed region using sharkfin diagram in the $a-M$ plane for
different values of $L_{AdS}$. The regions with $\cal D \gsim$$0$ corresponds to physically allowed non-extremal 
black hole configurations. The boundaries are the extremal configurations. Comparing to Fig.\ref{sharkdS} 
for dS cases, it is seen that the allowed region in AdS background in Fig.\ref{sharkAdS} 
is generally larger. The attractive nature of the negative cosmological constant in the AdS case  
satisfy the horizon formation criterion in much larger region than the dS cases. We have also understood 
from the bottom-right panel of Fig.\ref{sharkAdS} that for larger values of the $L_{AdS}$ the 
effect of cosmological constant weakens --  hence the allowed regions are shrinked and eventually  
leads to the standard Kerr limit.

\begin{figure}[ht!]
\begin{center}
\raisebox{0.0cm}{\hbox{\includegraphics[angle=0,scale=0.24]{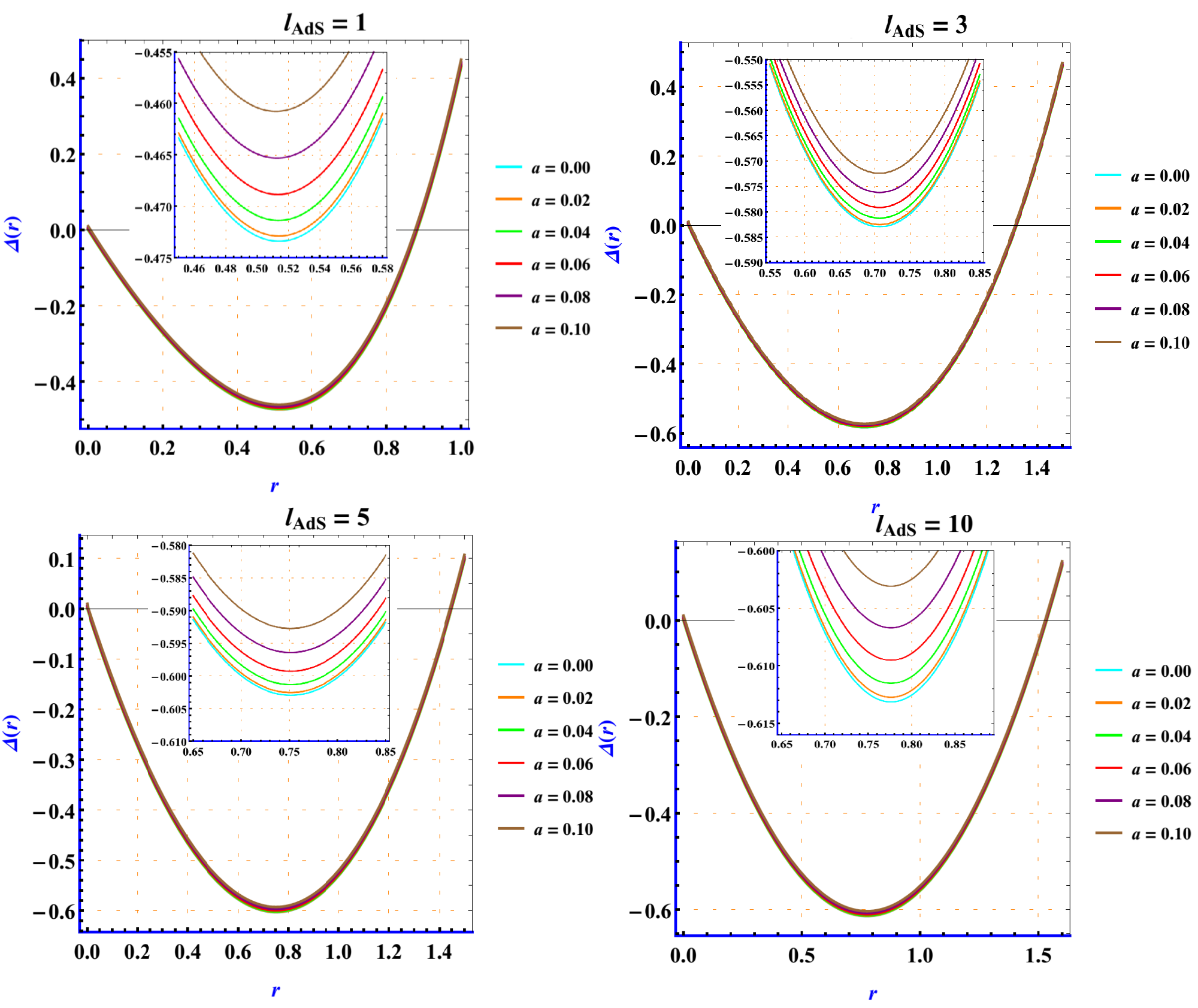}}}
\caption{The variation of the discriminant $\Delta(r)$ term with radial distance ($r$)
	for different values of $L_{AdS}$=$1,3,5$ and $10$ and spin parameter 
	of the Kerr-BH with $a$ = $0.00$, $0.05$ and $0.10$.}
\label{discriAdS}
\end{center}
\end{figure}

In Fig.\ref{discriAdS} we have plotted the variation of the horizon function $\Delta(r)$  
as a function of radial coordinate($r$) for different values of the spin parameter $a$ 
and $L_{AdS}$. As discussed before that the horizon radii are corresponds to $\Delta(r)=0$. 
From all the plots in Fig.\ref{discriAdS} it is clear that all the curves intersect the radial axis at two points -- 
the inner (Cauchy) and the outer (event) horizons. Thus the Kerr-BH with NLED-AdS model always 
has two distinct boundaries. The values of the rotation parameter shift the roots slightly. For example, 
with $a=0.10$ the Cauchy(event) horizons shift outwardly(inwardly). With lower values of $a$ the effects 
of these shifting is negligible. For larger values of $L_{AdS}$, for example  
the bottom right panel of Fig.\ref{discriAdS}, the effect of cosmological constants diminishes 
and BH behaves as standard Kerr-BH in the NLED model. By comparing Fig.\ref{discridS} in the dS cases, 
we see the absence of cosmological horizon in AdS. This is due to the 
attractive nature (equivalent to the presence of gravitating mass because of the negative cosmological 
constants) of the AdS background which has definite confinement in contrast to the expanding 
dS background (which mimics repulsive gravity).

\begin{table}[t!]
\centering
{\scriptsize
\begin{tabular}{||c|c||c|c|c|c|c|c|c|c|c|c||}
\hline
\hline
$a$&$L$& $r_{-}$ & $r_{+}$ & $r_{C}$ & $T_{r_{+}}$ & $T_{r_{C}}$ & $\Omega_{r_{+}}$ & $\Omega_{r_{C}}$ &$S_{r_{+}}$& $S_{r_{C}}$\\
\hline
\hline
0.00&5&0 &1.81&3.85 & 0.0268& 0.0160&0.0&0.0&10.25& 46.46\\
0.00&10&0 &1.61&9.09&0.0301 &0.0998&0.0&0.0& 8.177& 259.91\\
\hline
0.05&5&0.0016 &1.80&3.85 &0.0268 &0.0160&0.0153&0.0039&10.24& 46.47\\
0.05&10&0.0016 &1.61&9.09 &0.0455&0.0130&0.0192&0.0006& 8.162& 259.91\\
\hline
0.10&5&0.00639 &1.798&3.85 & 0.0268 &0.0161&0.0308& 0.0068& 10.19& 46.50\\
0.10&10&0.00640 &1.606&9.10 & 0.0453 &0.0129&0.0386&0.0012& 8.135& 259.92\\
\hline
\hline
\end{tabular}
}
\caption{The horizons radius, temperature, angular velocity and entropy 
	for different spin-parameter of the Kerr-BH for $L_{dS}=5$ and $10$.}
\label{tab:thermodS}
\end{table}

We have tabulated the horizons radius, temperature, angular velocity and entropy{\footnote{In \cite{Ravuri:2024fva} 
the authors shown  the solutions of the differential equation and get the thermodynamics parameters, i.e, 
the surface gravity and entropy for the whole regions from the AdS boundary to the inner horizon. In our 
analysis we have not done numerical solutions from the dS length to the inner horizon. Thus we 
estimated the temperature (i.e., the surface gravity) angular velocity, and entropy in the vicinity of 
the horizons only.}} for different spin-parameter of the Kerr-BH in Table.\ref{tab:thermodS}.

As we have considered the angular velocity of the BH is much smaller, i.e., $a \lsim 0.1$, thus 
squaring the spin-parameter is of the $\cal O$$(-2)$ thus the numerator is not changing appreciable 
in the expression of entropy in Eqn.\ref{entropy}, while 
the denominator is somehow appreciable affect together with $\Lambda$. 
The sign of the cosmological constant has also some affect due to its presence in 
the the denominator of Eqn.\ref{entropy} in dS and AdS background. It is clear from the 
Eqn.\ref{angular} that for fixed values of $a$ the large entropy corresponds to the 
low values of the angular momentum.

In Table\ref{tab:thermodS} we showed numerical solutions for the horizon radii and corresponding 
thermodynamical quantities for the Kerr-BH in a dS background. Under the non-rotating limit 
($a=0$) the inner horizon vanishes as it is same as the singularity in the Schwarzschild case. 
While, for finite rotation it arises as a small but non-zero value indicating its purely 
rotational origin. The event horizon remains of the order of unity and shows only weak 
dependency on $L_{dS}$. The cosmological horizon is correlated with the $L_{dS}$ and for large $L_{dS}$ 
the repulsion is weaker and the values are more close the $L_{dS}=10$ in contrast to $L_{dS}=5$.
The temperature differences between the event horizons and the cosmological horizon mimics the presence 
of two different thermal scales and hence the non-equilibrium thermodynamic is crucial to study further 
for Kerr-BH. The angular velocity vanishes in the static limit. We see from Eqn.\ref{entropy} that 
the entropy depends on the area, i.e., to say that for larger $r$ the thermodynamic entropy is large.
If we look more closely the expression of angular momentum in Eqn\ref{angular} we see that
for somewhat same values of the entropy, the angular velocity is proportional to $a$. This 
reflects the suppression of the rotational effects at large distances. Overall, this 
table suggests that the thermodynamic parameters are having intricate dependencies among $M(r)$, 
$a$ and $L_{dS}$.

\begin{table}[t!]
\centering
{\scriptsize
\begin{tabular}{||c|c||c|c|c|c|c|c|c|c|c||}
\hline
\hline
$a$&$L$& $r_{-}$ & $r_{+}$ &$T_{-}$ & $T_{+}$  &$\Omega_{-}$ & $\Omega_{+}$& $S_{r_{+}}$& $S_{r_{C}}$\\
\hline
\hline
0.00& 1&0 &0.88280 &{\it img.}&0.031&0.0&0.0&0.0&2.4484\\
0.00& 3&0 &1.31699 &{\it img.}&0.095&0.0&0.0&0.0&5.4490\\
0.00& 5&0 &1.44908 &{\it img.}&0.095&0.0&0.0&0.0&6.5968\\
0.00&10&0 &1.53465 &{\it img.}&0.055&0.0&0.0&0.0&7.3990\\
\hline
\hline
0.05& 1&0.00159 &0.88129 &49.85&0.299& 19.93&0.064& 0.00788 & 2.4540\\
0.05& 3&0.00159 &1.31555 &49.85&0.952&19.97&0.025& 0.00786 & 5.4465\\
0.05& 5&0.00159 &1.44759 &49.85&0.069&19.98&0.024& 0.00786  & 6.5918\\
0.05&10&0.00159 &1.53309 &49.84&0.055&19.98&0.021& 0.00786 & 7.3920\\
\hline
\hline
0.10& 1&0.00640&0.87673 &12.35&0.30&9.86&0.130& 0.03186& 2.4709\\
0.10& 3&0.00639&1.31220 &12.35&0.095&9.95&0.058& 0.03158& 5.4388\\
0.10& 5&0.00639&1.44308 &12.34&0.068&9.96&0.048& 0.03156& 6.5764\\
0.10&10&0.00639&1.52839 &12.35&0.056&9.97&0.042& 0.03155& 7.3709\\
\hline
\hline
\end{tabular}
}
\caption{The horizons radius, temperature, angular velocity and entropy 
	for different spin-parameter of the Kerr-BH for $L_{AdS}=1,3,5,10$.}
\label{tab:thermoAdS}
\end{table}

In Table\ref{tab:thermoAdS} we review the numerics of thermodynamic parameters 
for the AdS background. In this scenario we have only two horizons, namely inner(Cauchy) and event horizons 
due to the attractive nature if the negative $\Lambda$ with cosmological horizon being absent. In 
non-rotating limit, the inner horizon collapse to zero and associated temperature becomes 
unphysical indicating the absence of the physically meaningful inner thermodynamical structure. 
However, for rotating Kerr-BH, the inner horizon reappears with a finite radius and it is 
characterized by the extremely large temperature satisfying $T_{-}>>T_{+}$. 
The values of the angular velocities also clearly show the difference 
as $\Omega_{-}>>\Omega_{+}$ following Eqn.\ref{angular} as large entropy 
leads to low values of $\Omega$ and vice-versa. Unlike the dS case, the thermodynamic 
behavior is dominated by the event horizon in AdS. The absence of the cosmological 
horizon suggest the confining nature of the AdS background. 

\section{Summary and Outlook}
\label{sec:summary}

The existence of the BH is naturally emerges from the metric solution of the
GTR. The classical BHs are characterized by its mass (M), angular momentum
(L) and charge (Q). The simplest BH is characterized by its mass only (called as
Schwarzschild BH) and the metric have two singularities: one physical ($r=0$) and
one co-ordinate ($r=2M$ is called as event horizon). The addition of angular 
momentum together with cosmological constants in the background of NLED leads 
to very different characteristic features of the thermo-dynamical 
parameters of the Kerr-BH in dS and AdS background.

With NLED, the effective mass of the BH becomes a distribution M(r) and shown the mass profile 
of the Kerr-BH with different combinations of $q_m$ and  $\beta$ . This mass profile 
demonstrates that NLED replaces the singular core (like Schwarzschild cases) of the BH 
with a finite, smooth energy distribution, where the magnetic charge $q_m$ controls the 
total mass scale, while the nonlinearity parameter $\beta$ regulates the concentration of energy  
near the core (larger $\beta$ leads to more suppression of the energy density with diffuse core), 
leading to a finite and regular BH configuration. This is consistent with the role of NLED 
in resolving singularities in regular black hole.

We have shown the allowed $a-M$ plane for the physical viable solutions of the Kerr-BH  with 
positive discriminant, $\cal D \gsim$ $0$ using sharkfin diagrams. The contrasting features 
of dS (in Fig.3) and AdS (in Fig.5) are clearly visible, i.e., in dS case 
due to the positive $\Lambda$ (repulsive in nature) the allowed $a-M$ parameter 
space is more restricted than the AdS case because of the negative $\Lambda$ (effectively 
enhances the gravitational confinement) for the same values 
of the cosmological length. In both the cases the parameter spaces approaches 
the asymptotically flat Kerr limit for $L=10$.

We tabulated the horizon radii and thermodynamics parameters in dS and AdS scenario. For small 
rotation $a \lesssim 0.1$ entropy is mainly influenced by $\Lambda$. Larger entropy 
corresponds to lower angular velocity. In dS the inner horizon arises purely due to rotation and 
the cosmological horizon depends strongly on $L_{dS}$. The presence of multiple horizons 
with different temperatures highlights non-equilibrium thermodynamics. In AdS background only 
inner and event horizons exist; the inner horizon vanishes in the non-rotating limit but 
reappears with very high temperature and angular velocity with $T_{-} >> T_{+}$.

Our results highlight the intricate interplay between rotation, NLED effects and the 
cosmological constant for the numerical values of the horizon radii 
and thermodynamics parameters. We will report elsewhere the 
analysis beyond the slow-rotation approximation.  The presence of magnetic charge introduces 
an addition thermodynamic degree of freedom,i.e., the magnetic potential. Furthermore, the 
cosmological constant mimics the thermodynamic parameter pressure following $P=-\frac{\Lambda}{8\pi}$.  
Therefore the thermodynamic stability check and understanding the phase structure are 
of paramount interest. The observational signatures of the geodesic motion of a particle 
approaching the BH and stability of the orbit, the shadow structure, and gravitational 
wave emission from the Kerr-BHs with NLED corrections in dS and AdS background are 
important avenue to explore further.

\subsubsection*{Acknowledgments}  

We acknowledge the High Performance Computing (HPC) facility of the Shivaji
University Kolhapur for our numerical simulation.

\end{document}